# Use of two Public Distributed Ledgers to track the money of an economy

Name: Gonzalo         Family name: Garcia-Atance Fatjo

ggarcia-atancefatjo@uclan.ac.uk

University of Central Lancashire, Preston PR1 2HE, UK

https://orcid.org/0000-0002-3914-7160

**Abstract**

A tool to improve the effectiveness and the efficiency of public spending is proposed here. In the 19$^{th}$ century banknotes had a serial number. However, in modern days the use of digital transactions that do not use physical currency has opened the possibility to digitally track almost each cent of the economy. In this article a serial number or tracking number for each cent, pence or any other monetary unit of the economy is proposed. Then, almost all cents can be tracked by recording the transactions in a public distributed ledger, rather than recording the amount of the transaction, the information recorded in the block of the transaction is the actual serial number or tracking number for each cent that changes ownership. In order to keep the privacy of the transaction, only generic identification of private companies and individuals are recorded along with generic information about the concept of transaction, the region and the date/time. A secondary public distributed ledger whose blocks are identified by a hash reference that is recorded in the bank statement available to the payer and the payee allows for checking the accuracy of the first public distributed ledger by comparing the transactions made in one day, one region and one type of concept. However, the transactions made or received by the government are recorded with a much higher level of detail in the first ledger and a higher level of disclosure in the second ledger. The result is a tool that is able to accurately track public spending, to keep privacy of individuals and companies and to make statistical analysis and experiments or real tests in the economy of a country. This tool has the potential to assist public policymakers in demonstrating the societal benefits resulting from their policies, thereby enabling more informed decision-making for future policy endeavours.

**Graphical abstract**

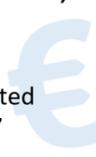

**How it works:**

It is composed of two, or more, permissioned public blockchains or public distributed ledgers that work in layers of detail and identification of parties. In the version of two ledgers. The first ledger records the movement of each cent in every transaction and has the highest level of detail but also the highest level of anonymity. The secondary ledger has the means for each party to check that their transaction is properly recorded but it has not details that might allow the tracking of the money up or down. The first layer allows for full tracking of the money up and down in the time.

Alternatively a version of three ledgers could also be implemented to protect the privacy at the same time as it registers higher level of detail. The first ledger records each cent but the level of details is limited, the second ledger has a high level of detail but it is lacking the tracking number of the cents, the third ledgers allows for checking the recording of the transaction in the other two ledgers by parties of the transaction as they have access to the reference of the block in the third ledger. The possible advantage of the three ledger system is that for example, an individual that happens to be the only one that paid for a type of service on one date and one region will not be able to track up or down the actual cents, keeping the privacy of the other parties. Other combinations of number of ledgers splitting the information can be thought of as a way to increase detail and keeping privacy. For example, a ledger with its own tracking numbers with details of place and date, another ledger with its own tracking numbers with details of concept and date and a third ledger for checking up the other two.

An alternative or complementary way to keep privacy is that although all cents are tracked, not all cents carry the same amount of details, for example, when the government is not involved a random percentage between 0-50% of the cents carry the details of the transaction (the concept, the date/time, the region) while the rest only carry a more general concept, the date/time and the region. In this way, statistical analysis and experiments are possible while making impossible for individuals to track up or down their payments with certainty.

Another way to improve privacy is to increase the size of the regions and also to record the time with less detail, such as days instead of minutes or weeks instead of days.

The banks and other financial institutions, including the central bank will have the duty, by law, to record all transactions in the two, or more, ledgers. They will also have a separate and private data base of the actual tracking numbers of each cent that are in the balance of the accounts. If they create new money, with a loan for example, they are responsible to assign a new tracking number to each cent. If the money is extinguished, they are responsible to eliminate the tracking number. This action will be recorded in the first ledger and in the secondary ledger. All transactions are recorded in both ledgers. In the first ledger the tracking numbers of each cent that is transferred with the transaction is recorded along with details of the transaction but with anonymised data. If the transaction is to, from or within the government the data is not completely anonymised and the level of detail is higher. For the secondary ledger, a hash reference is created for the transactions made to, from and within private companies or individuals to create a block with fewer details as date, region and type of concept, they will have access to the hash and being able to find the block with that hash. In this way, the individuals and companies involved in the transaction can check that their transaction was actually recorded as the total amount for one day, one type of concept and one region should coincide in the first ledger and the secondary ledger. The data use from the first ledger for hashing could be limited, less detailed, or include random number in order to protect from a brute force attack.

For a country in the euro area, it can implement this tool independently with the following criteria, if the deposit is in a bank within the country, all cents in the deposit should have the tracking number or serial number, all money that goes to another country loses the tracking number, all money that goes into the country gets a new tracking number. The tracking numbers are only for information purposes but not for accounting, the Public Distributed Ledger is not the actual ledger of the banks and it does not take over its place.

**Integration with banking system and central bank.**

Although briefly commented already, here, I am putting forward some of the ways this tool could work with the banking system. The public distributed ledgers would be maintained by the financial institutions that make transactions of money and payments within the economy. They should have a recording of the tracking number of each cent that is deposit by their customers in a database that is private and shared with the central bank to avoid duplicates. When a loan is given, they will be responsible or assigning new tracking numbers. When the loan is repaid, they will also be responsible for eliminating the tracking numbers of the money that is extinguished.

In the fractional reserve banking, part of the money from the deposits is actually deposited in the central bank of the country. This represents a challenge for this tool, as the tracking number of the cent can be held only in one place at a time. A solution could be that in the same way as the bank gives money to a borrower, and for that, it needs to create the tracking numbers of the cents, it gives money to the central bank and for that it creates also the tracking numbers of those cents. The result is that the loans and the reserves are assets against the deposits that are the liabilities. The tracking numbers held in the reserve might not be counted for the monetary base.

When a deposit is made in a bank account using a banknote, a doble transaction would be recorded, one transaction to an operating account to input the actual tracking number cents of the banknote reclaimed from the central bank, and the second part of the doble transaction would be to move from the operating account tracking numbers of cents not linked to banknotes into the customers bank account.

When a withdrawal of a banknote is made, another doble transaction would be recorded, the tracking numbers from the bank account move to the operating account, and from the operating account the tracking numbers cents of the banknote would be returned to the central bank. The tracking number of cents related to a banknote would be identified for example by starting with BA as in the example later in this document.

The creation of tracking numbers in the event of money creation is something that needs to be carefully implemented. They should not follow a pattern, such as being consecutive, in order to prevent the identification of parties and their type of transactions, up or down the money flow.

Other situations that might need careful implementation is the event of a bank run. As the customers of the bank do not have access to the actual tracking numbers of the cents they have in deposit, it is not expected that the depositors might have a different approach or understanding of bank runs as they already have till now. In a bank run, although the tracking numbers of the cents in deposits exist, the bank will not have liquidity to face the situation. It is important to remark that the tracking numbers of the cents are just a tool to track the money but it is not the money itself. Using the similitude of a serial number of a car and the plate number of the car, I think it is better to use the term "tracking number" rather than "serial number" as it is well understood by the public that a car must have only one serial number but it might change plate number or even have two plate numbers at the same time, for example, when a car dealer uses a temporary plate number.

The central bank might have a pool of tracking numbers for new cents that are available when new money is created as well as a sink of tracking numbers for money extinguishment.

Finally, there are three options for the financial institutions when they need to choose what tracking numbers held in the bank account are going to be used in the transaction. 1) randomly selected, 2) first in- firs out and 3) last in – first out. Possibly, last in- first out represents better how the money moves and what is the level of savings within the economy. On the other hand, randomly selected will possibly cast the best statistical analysis with minimum amount of transactions although it will require more computational power to track the money as it will "dilute" quickly. However, this topic needs farther analysis.

**Statistical analysis, experiments and impact**

With the tool implemented, data will be publicly available to do statistical analysis of how the money moves within the economy and when and where the money is created. New enterprises will create software that processes the information for the public in a friendly manner, for free and/or as a paid service. It will also enable the establishment of organizations capable of correlating money transactions with the actual value of goods and services, thereby enhancing understanding of statistical analysis. These organizations will provide clearer insights into the relationship between money transactions and real-world economic value. It will also contribute to the spread of the understanding of the economy. It will show new market opportunities within the economy reducing the advantage that big actors have. It will cast information about the impact of public spending. Furthermore, it will facilitate daily monitoring of the monetary base and velocity of money with a high degree of precision, allowing for more accurate inflation estimation.

Experiments, or real tests, about how the economy works and about what spendings and taxes to increase or reduce from the point of view of public policy would e possible now that there would be an instant measurement of how the money moves within the economy. If every monetary unit in the economy can be monitored all the time, it could be possible to build experiments in order to optimize the economy. As an example one possible experiment would be that the VAT in some goods changes from 20% to 19.9% in all even months and to 20.1% in all odd months. The average is still 20% but this change could be high enough to introduce a signal measured by the blockchain, or public distributed ledger, tool and low enough to not modify the decision making of the consumers. This signal with a frequency of 2 months could be tracked down in the economy in different sectors and identify the impact of a change in the economy. Similar to radiomagnetic frequencies in a full of radiomagnetic noisy environment. Another experiment could be done with the payment and service provided by public servants. The state could make a group of public servant work 30 hours in even weeks and 40 hours in odd weeks and pay the public servant weekly. The signal and impact of the public servant in the economy could be measured. Other experiments more sophisticated and better thought through with other types of taxes and spending could be used inspired by frequency hopping in radio waves…

The tool could facilitate the identification of public spending that creates wealth and value as well as public spending that waste resources. In this way, the tool will significantly improve the efficiency and the effectiveness of public spending. Being the public spending in most countries between 30%-60% of the GDP [1] and being the taxes compulsory, the government bears the responsibility of ensuring that public spending is allocated to areas where it is most needed and where it can yield the greatest impact. Additionally, it may consider reducing taxes in cases where the intended impact of the spending falls short of expectations. Undoubtedly, most politicians would welcome the opportunity to utilize this tool to enhance their decision-making processes. The potential improvement in public spending, ensuring funds are allocated efficiently rather than wasted, could yield significant long-term benefits. Over time, this could lead to increased national wealth or potentially reduce the need for citizens to work as much as they do now to maintain their current standards of living. While the long-term impact of enhancing public spending and what proportion of the public spending can be improved are beyond the scope of this work, it's conceivable that there would be certain countries were around 2% of GDP, for example, could be saved from being wasted. That would have a substantial cumulative effect over several decades.

**Approximate calculations and example of blocks.**

The monetary base and the number of tracking numbers needed for the smallest possible monetary unit in different countries is given in table 1.

Table 1: Monetary base M2 retrieved from www.tradingeconomics.com [2] and number of tracking numbers needed.

| Country | Monetary base M2 | Currency | Tracking numbers =Cents x 100 | **Hexadecimal** |
|---|---|---|---|---|
| United Kingdom | 3 009 816 | GBP Million | 30 098 160 000 000 000 | **6AEE1DF7316000** |
| United States | 20 784 | USD Billion | 207 840 000 000 000 000 | **2E26560FA1E0000** |
| Euro area | 15 105 923 | EUR Million | 151 059 230 000 000 000 | **218AB91C031AC00** |
| China | 304 795 | CNY Billion | 3 047 950 000 000 000 000 =fens x 100 | **2A4C7E605326E000** |

The calculation is made assuming there are available 100 times more tracking numbers than cents in the monetary base M2.

Here, some quick calculations are provided in order to estimate the storage of data needed.

- How much data is a payment of **3000 euros**?
  Each cent is identified with 16 hexadecimal digits. That is 8 bytes.
  3000 euros are 300 000 cents, it needs 300 000 x 8 = 2 400 000 bytes ≈ **2.4 MB**
- A payment of **3 000 000 euros** would be **2.4 GB**
- The **GDP of euro area** [3] in bytes would be 14 210 000 000 000 x 100 x 8 ≈ **11 400 TB,**
- For the euro area, assuming a **cost of hard drive** per TB of $20, to store the whole track of each cent would be of the order of **$228 000 per year per node.**

Here, some example blocks (blocks of data to be added to the Public Distributed Ledger) are provided. The first ledger is in blue, it has the tracking numbers of each cent, and second ledger is in green, it has the reference that appears in the bank statements. Only data that is in bold font is actually recorded.

Blue blocks from first ledger can actually be divided into smaller blocks to improve privacy, so there is no identification of transactions based on the amount. Then green block reference would be a hash of the combination of the smaller blue blocks.

Government to private

| | |
|---|---|
| Reference: **535D3D0C**<br>Date: **2024/03/29   14:47 UTC**<br>Amount:  **1812.24 €**<br>Cent tracking numbers:<br>1-              **2A1F3B4E5CAB6D7F**<br>2-              **8E9A2C593B4D5F63**<br>…              …<br>181 223-    **B1C25A6B527D3E4F**<br>181 224-    **C3D46A7B81712E5F** | Payer:  **GOV- B4E5**   (Madrid/ Alcalá de Henares city Hall/ Accounting Services/ Human Resources)<br>Payee:  **2C3B4**  (Public servant generic identity)<br>Detailed concept: **s236** (Keeping accounting books)<br>Concept type: **w24** (Wages)<br>Region: **rA12** (Madrid East) |
| Reference: **22C5B3A7**<br>Date: **2024/03/29   14:48 UTC**<br>Amount:  **400.12 €**<br>Cent tracking numbers:<br>181 225-    **3B4E5C6D7F437D5F**<br>181 226-    **C3B3652A1F8E9A26**<br>…              …<br>221 235-    **D3E4F5AA66B5B1C2**<br>221 236-    **3DC4E5F6A7B83612** | Payer:  **GOV- B4E5**   (Madrid/ Alcalá de Henares city Hall/ Accounting Services/ Human Resources)<br>Payee:  **2C3B4**  (Public servant generic identity)<br>Detailed concept: **s236** (Keeping accounting books)<br>Concept type: **w24** (Wages)<br>Region: **rA12** (Madrid East) |
| *Reference hashed: **A46FE3923CB4561B0** (This is a hash from block 535D3D0C and 22C5B3A7 or its subdivisions, it appears in the bank statement)*<br>*Date: **2024/03/29***<br>*Amount: **2212.36 €*** | *Payer: **GOV- B4E5**   (Madrid/ Alcalá de Henares city Hall/ Accounting Services/ Human Resources)*<br>*Payee: **2C*****  (Public servant generic identity, partially covered)*<br>*Concept: **s23** (Administration)* |

Private to private

| | |
|---|---|
| Reference for blockchain :**2D33455D**<br>Date: **2024/04/18   10:56 UTC**<br>Amount:  **25 €**<br>Cent tracking numbers:<br>1-              **C3D4E5F6A737B812**<br>2-              **B1C252D3E4F5A6B5**<br>…              …<br>2499-        **5A6B5D3E4FA6B1C2**<br>2500-        **6A7B81C3D4C2E5F2** | Payer:  **AAAA1**  (Untraceable physical person, shared code)<br>Payee: **63B4F**  (Unidentifiable company of repairs and auto services, shared code)<br>Detail concept: **r522** (Tyre repair)<br>Concept type: **t41** (Payment of service in shop)<br><br>Region: **rC02** (Barcelona North) |
| *Reference hashed: **B6539CBA985FFA145** (this is a hash from block 2D33455D or its subdivisions, it appears in the bank statement)*<br>*Date:**2024/04/18***<br>*Amount: **25 €*** | *Payer: **AAAA1**  (Untraceable physical person, shared code)*<br>*Payee: **63B4F**  (Unidentifiable company of repairs and auto services, shared code)*<br>*Concept: **r52** (Auto repair)* |

Private to government

| Reference for blockchain : **25D4FAE3**<br>Date: **2024/06/28   10:12 UTC**<br>Amount: **25600.17 €**<br>Cent tracking numbers:<br>1-              **24E5F6A756C3DB81**<br>2-              **B1C294D3E4F5A6B5**<br>…              …<br>2560016-   **B1C2B5D3E2645A6F**<br>2560017-   **C3D144E5F26A7B81** | Payer:  **AAAA1**  (Untraceable physical person)<br>Payee: **GOV-AR3**  (Income Tax office bank account in Madrid)<br>Detailed concept: **t16** (Annual income tax 28%)<br>Concept type: **gr20** (Income tax paid by individual)<br><br>Region: **rC14** (Valencia) |
|---|---|

| *Reference hashed:* **FA146539CBA985CF5** *(this is a hash from block 25D4FAE3 or its subdivisions, it appears in the bank statement)*<br>*Date:***2024/06/28**<br>*Amount:* **25600.17 €** | *Payer:*  **AAAA1**  *(Untraceable physical person)*<br>*Payee:* **GOV-AR3**  *(Income Tax office bank account in Madrid)*<br>*Concept:* **t1** *(Annual income tax)* |
|---|---|

Government to government

| Reference for blockchain : **7E35D4FA**<br>Date: **2024/07/20   13:43 UTC**<br>Amount: **370632.11 €**<br>Cent tracking numbers:<br>1-              **24E5F6A7C243DB81**<br>2-              **E4F5A6BB121C2D31**<br>…              …<br>37063210-   **3E45A6F34B1C2B5D**<br>37063211-   **C3D4E512F26A7B81** | Payer:  **GOV-AR3**  (Income Tax office bank account in Madrid)<br>Payee: **GOV- B4E5**   (Madrid/ Alcalá de Henares city Hall/ Accounting Services/ Human Resources)<br>Detailed concept: **s22** (Budgeting human resources)<br>Concept type: **gr01** (Transfer according to budget)<br>Region: **rA12** (Madrid East) |
|---|---|

| *Reference un-hashed* **7E35D4FA** *(it appears in the bank statement)*<br>*Date:***2024/07/20**<br>*Amount:* **370632.11 €** | *Payer:*  **GOV-AR3**  *(Income Tax office bank account in Madrid)*<br>*Payee:* **GOV- B4E5**   *(Madrid/ Alcalá de Henares city Hall/ Accounting Serv./ Human Resources)*<br>*Concept:* **s22** *(Budgeting human resources)* |
|---|---|

Central bank reserve to printed money.

| | |
|---|---|
| Reference for blockchain: **FAE325A2**<br>Date: **2024/07/22   9:00 UTC**<br>Amount:  **10 €**<br>Cent tracking numbers:<br>1-              **F6A7C3DB2384E581**<br>2-              **A6B16B1C2D3E4F51**<br>…              **…**<br>999-          **36FB1C2B5E4245AD**<br>1000-        **E5F26A7B8C3D5641** | Payer:  **GOV-AA1** (Central bank reserve Spain)<br>Payee: **GOV-AA6B** (Printed money tracking number sink)<br>Detail concept: **X22868512967** (Serial number of physical banknote)<br>Concept type: **a15** (Money reprint)<br>Region: **rA1** (Madrid) |
| Reference for blockchain: **FAE325A4**<br>Date: **2024/07/22   9:01 UTC**<br>Amount:  **10 €**<br>Cent tracking numbers:<br>1-              **BA55311D0C70001**<br>2-              **BA55311D0C70002** (hex from 22868512967)<br>…              **…**<br>999-          **BA55311D0C70999**<br>1000-        **BA55311D0C71000** | Payer:  **GOV-AA6B** (Printed money tracking number sink)<br>Payee:  **GOV-AA6C** (Printed money tracking number source)<br>Detail concept: **X22868512967** (Serial number of physical banknote)<br>Concept type: **a14** (Money print)<br>Region: **rA1** (Madrid) |
| *Reference un-hashed **FAE325A2** (it appears in the bank statement)*<br>*Date:**2024/07/22***<br>*Amount: **10 €*** | *Payer:  **GOV-AA1** (Central bank Spain)*<br>*Payee: **GOV-AA6B** (Printed money tracking number sink)*<br>*Concept: **X22868512967** (Serial number of physical banknote)* |
| *Reference un-hashed **FAE325A4** (it appears in the bank statement)*<br>*Date:**2024/07/22***<br>*Amount: **10 €*** | *Payer:  **GOV-AA6B** (Printed money tracking number sink)*<br>*Payee: **GOV-AA6C** (Printed money tracking number source)*<br>*Concept: **X22868512967** (Serial number of physical banknote)* |

The public, individuals and companies, can check that their transactions are recorded using the reference in their bank statements. Following the example given, the total amount of payments to companies of repairs and auto services made on the 2024/04/18 must coincide in both blockchains. The company can check that the block *B6539CBA985FFA145* is in the secondary blockchain. Therefore it can make sure its transaction is recorded.

**Conclusion and future work:**

A tool to track all the digital money of an economy is presented here. This tool is for information purposes only and does not substitute any form of transferring money or any way of doing transactions. Several strategies to keep privacy are discussed. The tool presented here relies on these two ideas that are:

- To use a blockchain type technology so anybody can track every cent within the economy.
- To use two blockchains or public distributed ledgers to balance the conflicting constraints of recording details and privacy. While one blockchain records the anonymised details, the other one allows the individuals to check accurate recording of their transactions without compromising privacy.

The potential of this idea to enhance the efficiency and effectiveness of public spending over the long term is substantial. Economic actors of any size would have better information to make decisions. Proper implementation could lead to significant economic growth over several decades, aided by the principle of compound interest. Effective communication is crucial for its adoption. While some politicians may resist, many would likely support it, especially if the benefits are emphasised and better governance enhanced.

Future envisaged steps could be:

1. Develop academic papers to enhance the idea, address any shortcomings, define integration with near-money assets, batch process transactions and high frequency trading, set the recorded level of detail that is beneficial for the society, test for privacy breaches, publicly discuss all possible ethical issues, guard against potential use for government surveillance and establish a structured framework for experimentation and statistical analysis.
2. Promote the idea through collaborations with YouTubers, influencers, and opinion leaders to generate public interest and demand.
3. A political party agrees to incorporate the idea into their political campaigns and drafts legislation specifying the financial institutions tasked with maintaining and updating the publicly distributed ledger.
4. That party is elected and implements the tool in their country.
5. The tool is successful producing positive outcomes, leading to increased public demand in other countries.
6. There is an extensive adoption of the tool in most democracies.
7. 

**References:**

Although this document is not a traditional research article but rather a disclosure of a proposed tool, and while it doesn't heavily draw from research articles, it's essential to acknowledge some foundational contributions. Satoshi Nakamoto's work has significantly influenced public awareness of blockchain technology, upon which this document relies [4]. Additionally, in the realm of economic principles, Juan Ramon Rallo's tireless teachings have profoundly impacted public understanding and created awareness about public spending and inflation [5]. Actually, the idea of this tool came to my mind while watching one of his videos.

[1] International Monetary Fund (IMF). (2024). Government Expenditure, Percent of GDP. Access date 17/04/2024. Retrieved from
https://www.imf.org/external/datamapper/exp@FPP/USA/FRA/JPN/GBR/ESP/ITA/ZAF/IND/SWE

[2] Trading Economics. (2024). Money Supply M2. Access date 17/04/2024. Retrieved from https://tradingeconomics.com/country-list/money-supply-m2

[3] World Bank. (2024). Gross Domestic Product (GDP) data for Euro Area Access date 17/04/2024. Retrieved from https://data.worldbank.org/indicator/NY.GDP.MKTP.CD?locations=XC

[4] Nakamoto, S. (2008). Bitcoin: A peer-to-peer electronic cash system. Retrieved from https://bitcoin.org/bitcoin.pdf

[5] Rallo, J. R. (2019). Liberalismo: Los 10 principios básicos del orden político liberal. Madrid, Spain: Deusto.